\documentclass{article}
\pdfoutput=1
\usepackage{cite}
\usepackage[utf8]{inputenc}		
\usepackage{xcolor,graphicx}				
\usepackage{pifont,amsmath,bm,amsfonts,dcolumn,cancel,upgreek} %
\usepackage{amssymb}  
\usepackage{chemarr}  

\DeclareSymbolFont{MathSL}{OT1}{cmr}{m}{sl}
\DeclareSymbolFontAlphabet{\mathsl}{MathSL}
\newcommand{\pnvec}[1]{{\bm #1}} 

\newcommand{\ssrsmallfont}[1]{\text{\fontsize{6pt}{7pt}\selectfont #1}}
\newcommand{\ssrsmall}[1]{_{\ssrsmallfont{#1}}}

\newcommand{\Ndop}{\Nqmopvec{d}} 
\newcommand{\Npprate}{{\beta}}  

\def\ssref#1{Sect.~\ref{ss:#1}}

\def\eref#1{Eq.~\ref{e:#1}}

\def\erefs#1{Eqs.~\ref{e:#1}}

\newcommand{\Nfret}{FRET}

\newcommand{\exv}[1]{{\bigl\langle #1\bigr\rangle}}
\newcommand{\exvscripts}[3]{\lower.95ex\hbox{${\scriptstyle #1}$}\exv{#2}_{#3}}
\newcommand{\exvbigscripts}[3]{\lower.95ex\hbox{$#1$}\exv{#2}_{\textstyle #3}}
\newcommand{\ket}[1]{\bigl|#1\bigr\rangle}\newcommand{\bra}[1]{\bigl\langle #1\bigr|}
\newcommand{\ketx}[2]{\bigl|#1\bigr\rangle_{\textstyle #2}}\newcommand{\brax}[2]{\lower.95ex\hbox{$#2$}\bigl\langle #1\bigr|}

\newcommand{\abvalu}[1]{\lvert #1\rvert } 

\newcommand{\Ndz}{D_0}\newcommand{\Ndst}{D_\star}\newcommand{\Naz}{A_0}\newcommand{\Nast}{A_\star}
\newcommand{\NhamOp}{{{\Nmatrix{H}}}} 
\newcommand{\NdensOp}{{{\Nqmopg{\uprho}}}} \newcommand{\NdensElt}{{\rho}}
\newcommand{\Nfretr}{\Npprate}
\newcommand{\beq}{\begin{equation}}
\newcommand{\eeq}[1]{\label{e:#1}\end{equation}}
\newcommand{\Nsubsystem}{{\mathfrak s}}\newcommand{\Nenviro}{{\mathfrak e}}
\newcommand{\inv}{^{\raise.15ex\hbox{${\scriptscriptstyle -}$}\kern-.05em 1}}
\newcommand{\Nmatrix}[1]{\ensuremath{\text{\sffamily #1}}} 
\newcommand{\Nqmopvec}[1]{\ensuremath{\text{\sffamily\bfseries #1}}} 
\newcommand{\Nqmopg}[1]{#1}                                          
\def\rmi{\mathrm i}
\newcommand{\half}{\tfrac12}
\newcommand{\ex}[1]{{\mathrm e}^{#1}}                 
\newcommand{\dd}{{\mathrm d}}
\newcommand{\Ntranspose}{^{\mathrm t}}

\newcommand{\Ntracex}[1]{\text{Tr}_{\textstyle #1}\,}
\newcommand{\sunit}{\ensuremath{\mathsf s}}

\begin{document}
\null\vfil\begin{center}{\LARGE The Role of Quantum Decoherence in FRET}\end{center}
\medskip\begin{center}{\LARGE
Philip C. Nelson\\ Department of Physics and Astronomy\\ University of Pennsylvania\\
Philadelphia PA USA}\end{center}
\bigskip\noindent \textit{Running head:} Role of Quantum Decoherence in FRET
\vfil\newpage

\begin{abstract}
Resonance energy transfer has become an indispensable experimental tool for single-molecule and single-cell biophysics. Its physical underpinnings, however, are subtle: It involves a discrete jump of excitation from one molecule to another, and so we regard it as a strongly quantum-mechanical process. And yet, its kinetics differ  from what many of us were taught about two-state quantum systems; quantum superpositions of the states do not seem to arise; and so on. Although J. R. Oppenheimer and T.~F\"orster navigated these subtleties successfully, it remains hard to find an elementary derivation in modern language. The key step involves acknowledging quantum decoherence. Appreciating that aspect can be helpful when we attempt to extend our understanding to situations where F\"orster's original analysis is not applicable.  

\end{abstract}
The phenomenon of nonradiative resonance energy transfer, often called ``F\"orster'' or ``fluorescence'' resonance energy transfer (FRET), has become a central tool in biophysical instrumentation \cite{Ahwan09a}, starting from the demonstration that it could be observed at the single-molecule level in room-temperature, aqueous solution \cite{Ha:1996p2103}. Today FRET-based sensors detect intra- and intermolecular motions, local chemical environment, and even mechanical forces \cite{Chang:2016ch}, both in vitro and even inside living cells, with high time resolution. 

FRET displays both classical and quantum aspects. When we discuss it, we often imagine an excitation {state} of a {fluorophore} as a discrete \emph{thing} that can be transferred from a donor to an acceptor fluorophore intact, as basketball players pass the ball. That picture never mentions the possibility of quantum-mechanical superposition states, in which an excitation is delocalized (simultaneously located on two different fluorophores), even though we know that two-state quantum systems generally do show that phenomenon. Also, most descriptions tacitly assume that  the transfer can be described by a fixed probability per unit time; that is, the transfer follows a rate equation with {first-order kinetics}. Again, however, two-state quantum mechanical transitions generally do not  behave in this classical way (see \ssref{FRET.itssot} below). 

In short, although we might have expected behavior intermediate between the classical and quantum regimes, instead \Nfret{} seems to involve {fluorophore}s simultaneously displaying \emph{strongly} quantum behavior (discrete energy levels), but also \emph{strongly} classical behavior (no superpositions, localized excitations, first-order rate equations). It is worthwhile to see how this is possible.

The original quantum derivations by Oppenheimer and F\"orster 
\cite{PhysRev.60.158,Afors48b,Arnold:1950p5010}\nocite{Afors48a} appealed to the ``golden rule'' of quantum mechanics
\cite{Dirac:2006p2859}. However, Physics textbooks present that result in a way that seems applicable only for  transitions to or from a continuum of states---not the discrete electron energy levels of donor and acceptor fluorophores. Despite that discrepancy, these early derivations gave accurate results because the ``golden rule'' approach implicitly incorporates the crucial feature of fast quantum decoherence \cite{Acleg06a,Bnitz06a}. The aim of the present note is to keep that feature in plain view, using terminology borrowed from quantum electronics, e.g.\
\cite{Bagra82a}. Although known to specialists, this approach is simple enough to deserve wider currency.

For historical details about the discovery of FRET, see \cite{Masters:2014p15115,Knox:2012ji,Acleg06a}. More details about this derivation are given in \cite{Bnels17a}.

\section{Isolated two-state system\label{ss:FRET.itssot}}
To focus on the key issues, we will not discuss the process that excites the {donor}, nor the eventual fluorescence of the {acceptor}, instead concentrating on the transfer of the excitation from one to the other. We will also make some simplifying assumptions:
\begin{itemize}\item  We suppose that only two electronic states of the {donor} are relevant: the ground state $\ket{\Ndz}$ and one excited state $\ket{\Ndst}$. Similarly, we consider only two {acceptor} states $\ket{\Naz}$ and $\ket{\Nast}$. We are particularly interested in transitions between  joint states of the form
\beq\ket{\mathsl1}=\ket{\Ndst\Naz},\quad\ket{\mathsl2}=\ket{\Ndz\Nast}
,\eeq{fret12}
whose energies are nearly equal (the resonance condition). Direct transitions between those two states, without any photon emission, are therefore compatible with energy conservation.
\item We will eventually define a ``decoherence time'' $T$, and assume that it is much shorter than the hopping time ($T\ll\Omega\inv$ below). We will also assume that $T$ is much shorter than the mean waiting time before loss processes other than \Nfret{} deexcite the donor ($T\ll\tau$ below).
\end{itemize}

The transition between the states in \eref{fret12} would be easy to describe in a world containing only two atoms \cite[chapts.~7--9]{BfeynIII}. Suppose for a moment that the two states of interest have exactly the same electronic-state energy in isolation (they are exactly resonant). Make the convenient convention that these energy values are $E_\mathsl{1}=E_\mathsl{2}=0$.  When the two atoms are brought near each other, they will have a coupling giving rise to a Hamiltonian operator with an off-diagonal entry in the $\mathsl1,\mathsl2$ basis, which we may take to be real:
\beq\NhamOp=\begin{bmatrix}0&V\cr V&0
\end{bmatrix}
.\eeq{hamsimple}
The system's evolving state can then be expanded as 
\beq\ket{\Psi(t)}=a(t)\ket{\mathsl1}+b(t)\ket{\mathsl2}, \eeq{tevolu}
where the coefficient functions obey the Schr\"odinger equation:
$$\rmi\hbar\begin{bmatrix}\dd a/\dd t\cr\dd b/\dd t\end{bmatrix}=V\begin{bmatrix}b\cr a\end{bmatrix}
.$$

Consider the solution with the initial state $\ket{\Psi(0)}=\ket{\mathsl1}$; at later times, we find that $\abvalu{b(t)}^2=\sin^2(\Omega t/2)$, where $\Omega=2V/\hbar$. Interpreting this quantity as the probability to find the system in state $\mathsl2$, we conclude that the probability initially increases with time as $t^2$. But this means that the \emph{initial} growth rate of the probability is zero, contrary to the first-order kinetics observed in FRET. Moreover, at almost every time the state is a quantum superposition of $\ket{\mathsl1}$ and $\ket{\mathsl2}$, in contrast to the ``basketball'' picture of resonance energy transfer alluded to above. Finally, the solution just found is oscillatory: The system periodically reverts to being completely in state \textsl1, in contrast to the one-way transfer characteristic of \Nfret.

\section{Two-state system with environment}
\subsection{\label{ss:FRET.eembt}}
To see where we have gone astray, we must remember that our two {fluorophore}s are hardly alone: They are just a \emph{subsystem} of the entire world. Each constantly suffers collisions with surrounding water molecules, as well as less obvious influences involving fluctuating electric fields in its neighborhood and so on. A good {fluorophore} is robust to such disturbances, in the sense that they rarely knock it into a different electronic state. Nevertheless, environmental influences can affect the quantum-mechanical \emph{phase} of a fluorophore's state, by momentarily perturbing its energy levels during each collision. 
One way to incorporate this effect is by coupling both the donor and acceptor to a ``bath,'' for instance, of harmonic oscillators \cite{Jang:2007p10149,Bjang14a}. But the exact nature of the bath turns out not to be very important, leading us to suspect that there must be an approach that is not so explicit, and hence is computationally simpler.

A state measurement that could in principle be made internally to the donor--acceptor subsystem $\Nsubsystem$ corresponds to an observable $\Nmatrix O$ that acts only on the subsystem's two-dimensional state space ${\cal H}_{\textstyle\Nsubsystem}$. Given a pure state, we can express the measured value of such an observable without needing to know anything about the environment $\Nenviro$:
\beq\exv{\Nmatrix O}=\exvbigscripts{\Nsubsystem}{\psi|\Nmatrix O|\psi}{\Nsubsystem}
\qquad\text{for a pure state $\ket\Psi=\ketx\psi\Nsubsystem\otimes\ketx\phi\Nenviro$.}\eeq{exvpure}

Unfortunately, even if we could prepare a pure initial state, it would quickly evolve into an {entangled state} due to
the interactions between $\Nsubsystem$ and $\Nenviro$. However, we can still compactly summarize the effect of the environment on the measured values of observables that, like $\Nmatrix O$, refer only to the subsystem. To do this, we introduce a Hermitian operator $\NdensOp$ on ${\cal H}_{\textstyle\Nsubsystem}$ called the {density operator}, defined by constructing the dyad $\ket{\Psi}\bra{\Psi}$ and taking the trace over the environment state space \cite{Bschu10a,Bberm11a}:
\beq\NdensOp=\Ntracex\Nenviro\Bigl(\ket{\Psi}\bra{\Psi}\Bigr)
.\eeq{dfdensop}
In our problem, $\NdensOp$ can be represented by a two-dimensional matrix with respect to the basis $\ket{\mathsl1}$, $\ket{\mathsl2}$. 
If we know $\NdensOp$, then the measured value of any subsystem  observable can be expressed as 
\beq\exv{\Nmatrix O}=\Ntracex\Nsubsystem(\NdensOp\Nmatrix O)
\qquad\text{for any state, represented by $\NdensOp$.}\eeq{exvarb}

In order for this formulation to be useful, we need to be able to compute $\NdensOp$, at least approximately. This is not difficult when $\Nsubsystem$ is perfectly isolated from its environment, because in that case a pure (unentangled) state remains pure:
\beq\ket{\Psi(t)}=\ketx{\psi(t)}\Nsubsystem\otimes\ketx{\phi(t)}\Nenviro
\qquad\text{for isolated subsystem.}\eeq{evolisol}
Here $\ketx{\psi(t)}\Nsubsystem$ denotes the time development of the subsystem under its Hamiltonian, independent of that of  the environment, $\ketx{\phi(t)}\Nenviro$. 

Still restricting to the case of an isolated subsystem, the time development of $\NdensOp$ is determined by $\NhamOp_{\textstyle\Nsubsystem}$, the subsystem's Hamiltonian operator:
\beq\frac{\dd\NdensOp}{\dd t}=\frac1{\rmi\hbar}[\NhamOp_{\textstyle\Nsubsystem},\NdensOp]
\qquad\text{for isolated subsystem.}\eeq{densevol}
Notice that \erefs{tevolu}, \ref{e:dfdensop}, and \ref{e:evolisol} give
\beq\NdensOp(t)=\ketx{\psi(t)}\Nsubsystem\brax{\psi(t)}\Nsubsystem\quad \text{so}\quad 
\NdensElt_{ij}=\begin{bmatrix}
\abvalu{ a(t)}^2 & a(t)b(t)^*\cr a(t)^*b(t) & \abvalu{ b(t)}^2
\end{bmatrix}_{\!ij}
.\eeq{evolmatrixisol}
This formula shows that
the diagonal elements of $\NdensOp$ (``populations'') reflect the respective probabilities to be in the two states. Unlike the off-diagonal elements (``coherences''), the populations are unaffected if we change basis states to new versions differing by phases from the old ones, for example, $\ket{\mathsl1'}=\ex{\rmi\theta}\ket{\mathsl1}$.

\subsection{\label{ss:FRET.tedo}} 
As mentioned earlier, interactions with the environment $\Nenviro$ will destroy the simple form of \eref{evolisol},
converting an initially pure state to one that is entangled with the environment. Although these interactions are complicated, \ssref{FRET.eembt} above suggested that they could be summarized by saying that the subsystem's \emph{phase} is altered by the many environmental particles that interact with it. When we perform the trace operation in \eref{dfdensop}, the entanglement leads to the sum of many random phase factors in the off-diagonal elements of $\NdensOp$, effectively suppressing them within some
{decoherence time scale} $T$ \cite[chap.~3]{Bschl07a}. The diagonal terms are unaffected, however.

We must also extend the simplified discussion above (\eref{hamsimple}) by allowing for the possibility that the energies of $\ket{\mathsl1}$ and $\ket{\mathsl2}$ may not be exactly equal. Thus, let $\NhamOp=\NhamOp_0+\Nmatrix V$, where $\NhamOp_0$ is diagonal with eigenvalues $E_{\mathsl1}$ and  $E_{\mathsl2}$ and
$\Nmatrix V$ is the off-diagonal interaction operator appearing in \eref{hamsimple}.
\eref{densevol} then becomes
\begin{align}
\frac{\dd\NdensElt_{22}}{\dd t}&=\frac1{\rmi\hbar}[\Nmatrix V,\NdensOp]_{22}\label{e:ondevol22}\\
\frac{\dd\NdensElt_{ij}}{\dd t}&=\frac1{\rmi\hbar}\bigl([\Nmatrix V,\NdensOp]_{ij}+(E_i-E_j)\NdensElt_{ij}\bigr)
-\frac1T\NdensElt_{ij}\quad\text{for $i\not=j$.}
\label{e:offdevol}\end{align}
The environment enters via the last term above, which contains the
{decoherence time scale} $T$.

The donor can also lose its excitation directly, without transfer of energy to the {acceptor}. We approximate this effect as a decay term in the equation for $\NdensElt_{11}$:
\beq\frac{\dd\NdensElt_{11}}{\dd t}=\frac1{\rmi\hbar}[\Nmatrix V,\NdensOp]_{11}-\frac1\tau\NdensElt_{11}
.\eeq{densdiag}
(\eref{ondevol22} neglects the analogous effect for acceptor deexcitation, which is not relevant to our discussion.) \erefs{ondevol22}--\ref{e:densdiag} are sometimes called ``Pauli master equations,'' or ``Redfield equations'' \cite{Silbey:2011hd}.

\section{FRET\label{s:FRET.FRET}}
\subsection{Formal solution\label{ss:FRET.wccil}}
F\"orster studied the situation in which the decoherence rate, $1/T$, is much faster than either the transition rate, $\Omega=2V/\hbar$, or the donor deexcitation rate, $1/\tau$ (the ``fast decoherence'' limit). {Typical numbers for {chromophore}s in solution are $1/T\approx10^{14}\,\sunit\inv$ \cite{Gilmore:2008p4932}, compared with typical rates for donor fluorescence and mixing $\tau\inv\approx\Omega\approx10^8\,\sunit\inv$.}
F\"orster realized that in this situation, the environment effectively supplies a continuum of final states, even though the subsystem states of interest are discrete. This observation may have motivated him to apply the ``golden rule,'' initially developed for the emission of a photon into an explicit continuum of final states. Rather than appeal to this black box, however, we can equally well proceed simply by expansion in powers of $T\Omega$, as follows.

Let $S=(E_{\mathsl1}-E_{\mathsl2})/\hbar$.
Change variables to the four real quantities $U=\NdensElt_{11}$, $W=\NdensElt_{22}$, $X=(\NdensElt_{12}-\NdensElt_{21})/\rmi$, and $Y=\NdensElt_{12}+\NdensElt_{21}$. Then the dynamical equations take the real form 
\begin{equation}\begin{split}
	\dd U/\dd t&=-\half\Omega X- U/\tau \\
	\dd W/\dd t&=\half\Omega X\\
	\dd X/\dd t&=\Omega(U-W)-X/T-SY\\
	\dd Y/\dd t&=-Y/T+SX
.\end{split}\label{e:realfors}\end{equation}
This is a set of coupled linear differential equations with constant coefficients, so its solutions will be combinations of exponentials. 

Let $\pnvec Z(t)$ be the 4-component vector with entries $U(t)$, $W(t)$, $X(t)$, and $Y(t)$, so that \eref{realfors} can be written symbolically as $\dd{\pnvec Z}/\dd t=\Nmatrix M\pnvec Z$, where $\Nmatrix M$ is a $4\times4$ matrix.
When the coupling $\Omega=0$, we easily find one solution: 
\beq\pnvec Z_0(t)={\small\begin{bmatrix}U(t)\\ W(t)\\ X(t)\\ Y(t)\end{bmatrix}} = \ex{-\Nfretr_0 t}\pnvec B_0\quad\text{where}\quad
\pnvec B_0=
{\small
\begin{bmatrix}1\\ 0\\ 0\\ 0\end{bmatrix}}\quad\text{and}\quad \Nfretr_0=1/\tau.\eeq{zerothfret}
This solution describes spontaneous deexcitation of the {donor}, for example, via  fluorescence.

At small but nonzero $\Omega$, we expand all quantities in powers of $\epsilon=(T\Omega)$. For example, we expand the matrix $\Nmatrix M$ as $\Nmatrix M_0+\epsilon\Nmatrix M'$, with
$$\small \Nmatrix M_0 =  \left[ \begin{array}{cccc} -\tau^{-1} & 0 & 0 & 0 \\
            						       0 & 0 & 0 & 0 \\
						        	       0 & 0 & -T^{-1} & -S \\
							       0 & 0 & S & -T^{-1} \end{array} \right]
\text{\ and\ \,} \Nmatrix M' = \left[ \begin{array}{cccc} 0 & 0 & -T^{-1}/2 & 0 \\
            						       0 & 0 & T^{-1}/2 & 0 \\
						        	       T^{-1} & -T^{-1} & 0 & 0 \\
							       0 & 0 & 0 & 0 \end{array} \right].$$
Then we again seek a trial solution to \erefs{realfors} that is exponentially changing in time. Expanding the eigenvalue $\Nfretr$ as $\Nfretr_0+\epsilon\Nfretr'+\epsilon^2\Nfretr^{\prime\prime}+\cdots$, the first-order terms in the eigenvalue equation give $(\Nmatrix M_{0} + \Nfretr_0\mathbb I)\pnvec B' + (\Nmatrix M' + \Nfretr'\mathbb I)\pnvec B_0 = 0,$ while the second-order terms give $(\Nmatrix M_0 + \Nfretr_0\mathbb I)\pnvec B'' + (\Nmatrix M' + \Nfretr'\mathbb I)\pnvec B' + \Nfretr'' \pnvec B_0 = 0$.

Multiplying both sides of the first-order equation by the transpose $\pnvec B_0\Ntranspose$ and using \eref{zerothfret} gives $\Nfretr' = -\pnvec B_0\Ntranspose \Nmatrix M' \pnvec B_0$. Substituting the known $\pnvec B_0$ and $\Nmatrix M'$ gives $\Nfretr'=0$. At the next order, however,
\beq\Nfretr^{\prime\prime}=\half\frac{T\inv}{1+(TS)^2}
.\eeq{fretrate}
Altogether, we find  
\beq\Nfretr=\tau\inv + \frac{\Omega^2T/2}{1+(TS)^2}=\tau\inv + \frac{2V^2T}{\hbar^2+T^2(E_1-E_2)^2}
.\eeq{fr2}

\begin{figure}[h]
\begin{center}	\includegraphics{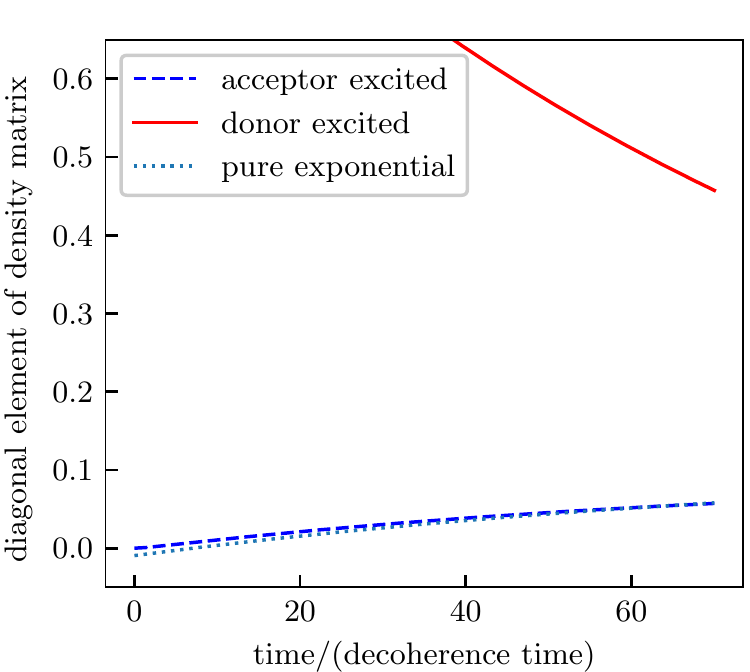}\end{center}
	\caption{\small Diagonal elements of the density matrix (``populations'') with sample parameter values $\Omega T=0.05$, $T/\tau=0.01$, and $ST=0.05$.
	These values are less extreme than the realistic ones mentioned in Sect.~\ref{ss:FRET.wccil}; they were chosen to make the crossover behavior visible in the graph. For comparison, the dotted line shows a single exponential that asymptotically matches the acceptor population at long times.}
\end{figure}%
\subsection{Numerical example}
To see the physical implication of \eref{fr2}, it is instructive to solve \erefs{realfors} numerically.
If we prepare the initial state $Z(0)=\pnvec B_0$, then strictly speaking, the initial rate of decrease of $\NdensElt_{11}$ is the first component of $\Nmatrix M\pnvec B_0
$, that is, $-1/\tau$. And the initial rate of increase of $\NdensElt_{22}$ is the second component, that is, 0. However, after a brief transient these behaviors change. Figure 1 shows
a numerical solution with sample parameter values. We see that,
although the time course of $\NdensElt_{22}$ is indeed initially flat, it soon starts to increase exponentially. Similarly, although $\NdensElt_{11}(t)$ initially starts to fall with slope $-1/\tau$, it soon starts to fall as $\ex{-\beta t}$, with $\beta$ given by \eref{fr2}. These effective first-order rate constants describe the excitation transfer.
The second contribution in \eref{fr2} shows that the transfer rate has a sharp maximum as a function of the energy difference. Importantly, the area under that peak does not depend on the value of the decoherence time $T$, as long as $T$ is small enough to justify the approximations made.

\subsection{Qualitative discussion}
We can now look back and identify the origin of the characteristic physical features of FRET. In general, irreversibility in classical or quantum physics stems from a system ``getting lost in phase space'' \cite{Bnitz06a}: The probability of returning to an initial state is vanishingly small. For a single atom emitting a photon into vacuum, irreversibility stems from the continuum of available photon states; in FRET, it stems from entanglement with the surroundings (decoherence).

In the limit of fast decoherence ($T\to0$), the third and fourth of Eqns.~\ref{e:realfors} say that $Y$ rapidly relaxes to zero, whereas the other coherence $X$ adiabatically tracks the quantity $T\Omega(U-W)$. Substituting that value for $X$ into the first two equations shows that, on the resonance $S=0$, the population difference $U-W$ has a contribution to its time dependence proportional to $T\Omega^2$, as we indeed found (\eref{fr2}).
In words, fast decoherence suppresses the effects of mixed quantum states, but one coherence is constantly  ``pumped up'' by the population difference, and feeds back negatively to it.

\subsection{Electric dipole  approximation\label{ss:FRET.ffada}}
To apply \eref{fr2} to \Nfret, we now recall that the interaction energy of two electric dipoles is proportional to the product of their electric dipole moments $\Ndop\ssrsmall D$ and $\Ndop\ssrsmall A$, and to the inverse cube of the distance between them. Specifically, in a molecular separation regime where dipole interactions dominate, $V$ in \eref{fr2} is proportional to 
$r^{-3}\exv{\mathsl2|\bigl(\Ndop\ssrsmall D\cdot\Ndop\ssrsmall A-3\Ndop\ssrsmall D\cdot\hat r\hat r\cdot\Ndop\ssrsmall A
\bigr)|\mathsl1}$, yielding the famous orientation dependence of the FRET rate \cite{Iqbal:2008p13773}. The rate is also proportional to $r^{-6}$, another key feature of FRET \cite{Sindbert:2011p5198,Wozniak:2008p5199}. (Note, however, that at very long distances our assumption of instantaneous (electrostatic) interaction fails. More detailed quantum-electrodynamics calculations show a gradual crossover to $r^{-2}$ behavior at large $r$, as we might have expected naively from the exchange of a real photon \cite{Bsala10a}.)

So far, we have assumed definite (exact) values for the donor's excited and ground state energies, and similarly for the acceptor. Actually, however, each of these energies changes over time due to molecular motions, that is, changes of the positions of the atomic nuclei. Accordingly,
we now introduce realistic (that is, broad) probability distributions of these energies, and average the mean rate for energy transfer over those distributions. The sharply peaked form of \eref{fr2} as a function of $E_1-E_2$ then implies that the mean \Nfret{} rate will be proportional to the overlap integral of the two distributions, another key feature of \Nfret. In fact, in the stated limit F\"orster was able to find a prediction with \emph{no free parameters} for the \Nfret{} rate, in terms of the  {donor}'s measured emission spectrum and fluorescence rate, the {acceptor}'s measured excitation spectrum and fluorescence cross section, the medium's {index of refraction}, and the distance and relative orientation between {donor} and acceptor.

In particular, the derivation just outlined explains a surprising aspect of \Nfret, which is that there can be highly specific energy transfer between two particular molecular species, despite the multitude of other directions into which the {donor} could instead emit a {photon}, and the crush of other molecules that could instead receive the energy:
\begin{itemize}\item To understand the dominance of \Nfret{} over photon emission, note that the ``near fields'' of a fluctuating dipole fall off with distance as $r^{-3}$, independent of its frequency. The ``radiation fields'' fall off more slowly, as $r\inv$, and they do depend on frequency. Turning these statements around, at \emph{small} distances the near fields are stronger by a factor of $(\lambda/r)^2$, where $\lambda$ is the wavelength of light corresponding to donor fluorescence. The square of this ratio can exceed $10^4$. 
	\item Turning to the other nearby molecules, the sharply peaked form of \eref{fr2} ensures that
only those with a transition resonant with the {donor}'s emission (overlapping spectra) will have significant probability per unit time to gain energy from it.\end{itemize}

\subsection{Limitations\label{ss:FRET.mrtrfp}}
Although the approximations made by Oppenheimer and F\"orster are often excellent for \Nfret{} used as a lab technique, later work has shown that they, and the physical picture that they support, do not hold in other situations, such as in photosynthetic apparatus.

First, we assumed that the interaction between two molecules can be approximated as a dipole-dipole interaction. For the  tightly spaced photosynthetic {chromophore}s, this approximation is not always valid. Not only does the multipole approximation break down, but ``exchange interactions'' (direct contact between the fluorophores' electron clouds) start to be significant---the Dexter mechanism.

Second, the derivation of \eref{fr2} assumed that excitation transfer is much slower than quantum {decoherence}. In that situation, we got a simple rate law for excitation transfer, and a nice picture of localized excitations. But there is a hierarchy of substructures in the photosynthetic apparatus, and within some of them the transfers are extremely fast. In this situation, it makes more sense to regard a whole array of {chromophore}s as a single ``supermolecule,'' with delocalized excitations called ``{Frenkel exciton}s.''  The supramolecular units in turn transfer excitons among themselves via \Nfret-like processes. For more about  these systems, see \cite{Jang:2013p10102,Strumpfer:2012p10868,Sener:2011p10860,Engel:2011p10775,Beljonne:2009p15144,Ishizaki:2009jg,Seibt:2017ey}.

\section{}FRET will continue to be the technique of choice for many more advances in single-molecule and single-cell biophysics, including some not yet imagined. This Perspective has pointed out that viewed as a physical phenomenon, FRET displays aspects not familiar from undergraduate (or even graduate) training in physics or chemistry, and these are key to its usefulness. Indeed, there were many missteps along the way to understanding FRET \cite{Masters:2014p15115,Knox:2012ji,Acleg06a}, but I have argued that an appreciation of the role of quantum decoherence leads directly to the main features.

\section*{Acknowledgments}
This work was partially supported by the United States 
National Science Foundation under Grant PHY--1601894. Some of the work was done at
the Aspen Center for Physics, which is supported by NSF grant PHY--1607611.

\bibliographystyle{unsrt}

\newpage
\begin{center}{\huge Caption}\end{center}
\begin{figure}[h]
	\caption{\small Diagonal elements of the density matrix (``populations'') with sample parameter values $\Omega T=0.05$, $T/\tau=0.01$, and $ST=0.05$.
	These values are less extreme than the realistic ones mentioned in Sect.~\ref{ss:FRET.wccil}; they were chosen to make the crossover behavior visible in the graph. For comparison, the dotted line shows a single exponential that asymptotically matches the acceptor population at long times.}
\end{figure}%

\end{document}